\documentclass[journal]{IEEEtran}
\usepackage{amsmath,amsfonts}
\usepackage{algorithmic}
\usepackage{algorithm}
\usepackage{array}
\usepackage[caption=false,font=footnotesize,labelfont=sf,textfont=sf]{subfig}
\usepackage{textcomp}
\usepackage{stfloats}
\usepackage{url}
\usepackage{verbatim}
\usepackage{graphicx}
\usepackage{amsmath}
\usepackage{amssymb}
\usepackage{adjustbox}
\usepackage{booktabs}
\usepackage{multirow}
\begin{document}

\title{Training-Efficient Text-to-Music Generation\\with State-Space Modeling}

\author{
\IEEEauthorblockN{
Wei-Jaw~Lee,
Fang-Chih~Hsieh,
Xuanjun~Chen,
Fang-Duo~Tsai,
and Yi-Hsuan~Yang\\}
\IEEEauthorblockA{
Graduate Institute of Communication Engineering, National Taiwan University}
\thanks{Correspondence to: Wei-Jaw Lee at f12942089@ntu.edu.tw.}
}



\maketitle

\begin{abstract}
Recent advances in text-to-music generation (TTM) have yielded high-quality results, but often at the cost of extensive compute and the use of large proprietary internal data. To improve the affordability and openness of TTM training, an open-source generative model backbone that is more training- and data-efficient is needed.
In this paper, we constrain the number of trainable parameters in the generative model to match that of the MusicGen-small benchmark (with about 300M parameters), and replace its Transformer backbone with the emerging class of state-space models (SSMs). Specifically, we explore different SSM variants for sequence modeling, and compare a single-stage SSM-based design with a decomposable two-stage SSM/diffusion hybrid design. All proposed models are trained from scratch on a purely public dataset comprising 457 hours of CC-licensed music, ensuring full openness.
Our experimental findings are three-fold. First, we show that SSMs exhibit superior training efficiency compared to the Transformer counterpart. Second, despite using only 9\% of the FLOPs and 2\% of the training data size compared to the MusicGen-small benchmark, our model achieves competitive performance in both objective metrics and subjective listening tests based on MusicCaps captions. Finally, our scaling-down experiment demonstrates that SSMs can maintain competitive performance relative to the Transformer baseline even at the same training budget (measured in iterations), when the model size is reduced to four times smaller.
To facilitate the democratization of TTM research, the processed captions, model checkpoints, and source code are available on GitHub via the project page: \url{https://lonian6.github.io/ssmttm/}.
\end{abstract}

\begin{IEEEkeywords}
Training Efficiency, State-Space Model, Text-to-Music Generation.
\end{IEEEkeywords}

\section{Introduction}
\IEEEPARstart{R}{ecent} text-to-music (TTM) generation models have exhibited impressive capabilities in generating high-quality audio from textual descriptions. Despite these advances, the practical deployment and academic adoption of TTM models remain severely constrained by issues of openness and reproducibility. In particular, existing systems often rely on proprietary training data and extensive computational resources, creating significant barriers for openness and reproducibility.
From the perspective of \emph{openness}, access to training data remains a major concern. A recent survey~\cite{batlle2025musgo} shows that only one out of 16 state-of-the-art TTM models (i.e., Stable Audio Open~\cite{evans2025stable}) is trained exclusively on publicly available, CC-licensed data. The remaining models rely on internal datasets or audio scraped from the Internet, which raises legal and ethical concerns.
\emph{Reproducibility} represents a significant practical challenge due to the substantial computational cost of training modern TTM models. Even for a relatively compact model such as MusicGen-small (300M parameters) \cite{copet2023simple}, training requires up to one million steps on 32 GPUs using approximately 20K hours of private audio data. Assuming training on a single consumer-grade GPU (e.g., RTX 3090) with BF16 precision, which is commonly available in academic environments, and following the heuristic in \cite{kaplan2020scaling}, the total training cost can be estimated to be on the order of $10^{20}$ FLOPs. This estimate is obtained as six times the product of the model parameter count and the total number of training tokens, where the latter is determined by the number of audio frames, batch size, and training steps. Under these assumptions, reproducing such a training pipeline would require over 4,044 GPU hours. Such computational requirements constitute a substantial barrier to faithful end-to-end reproduction, and become even more prohibitive when scaling to larger TTM models with billions of parameters.

Towards open and reproducible TTM research, in this paper we adopt MusicGen-small as the \emph{reference model}, and aim to build a new model that reaches compatible performance with MusicGen-small while satisfying the following two \emph{constraints}. 
First, like Stable Audio Open, we aim to use only \emph{publicly-available, CC-licensed data} to train our generation model from scratch (i.e., not fine-tuning). Specifically, we use a collection of around 450 hours of audio from Jamendo~\cite{bogdanov2019mtg}.
Second, to reflect realistic computational limitations faced by many academic researchers, all of our experiments are conducted on a \emph{single RTX 3090 GPU} with a restricted budget of at most \emph{100k training steps}.

Our reference model MusicGen-small employs 
a language model (LM) composed of 24 Transformer layers with self-attention, cross-attention, and feedforward modules. 
To attain comparable performance to MusicGen-small under the aforementioned constraints, the backbone of our new model has to be more data- and training-efficient than the Transformer.

Aiming to find a drop-in alternative of the Transformers, 
we explore using the emerging state-space models (SSMs), such as Mamba~\cite{gu2023mamba,dao2024mamba2} to build the LM of our TTM model.
SSMs can be thought of as linear recurrent neural networks (RNNs) with non-linear components. 
Unlike attention, which incurs quadratic complexity $O(n^2)$ with respect to sequence length, SSMs operate with linear complexity $O(n)$, making them more efficient for modeling long sequences. While attention enables each token to directly attend to all others—effectively functioning as a form of dynamic memory suited for retrieval-oriented tasks—SSMs maintain a fixed hidden state that is recurrently updated based on the input at each step. This design allows SSMs to compress and propagate information more compactly over time, enabling a stronger inductive bias for sequential understanding and improved computational efficiency, as demonstrated in many other domains.

To our best knowledge, little has been done to adopt SSMs for TTM. 
It remains unclear how to design an SSM-based LM to handle text conditions and to model the audio tokens.
In our study, we benchmark three SSM-based LM variants against a Transformer LM. Moreover, exploiting the coarse-to-fine structure of audio tokens, we adapt the two-stage framework of  DiscoDiff~\cite{lanzendorfer2025coarse} and use the proposed SSM-based LM for the coarse tokens, and
a pre-trained latent diffusion model 
for the finer-grained tokens. 
We compare the performance of the resulting model with that of the official MusicGen-small.

We aim to answer the following research questions (RQs):

\begin{itemize}

\item \textbf{RQ1}: Which SSM architecture better supports TTM, and how does it compare with the Transformers in terms of generation quality and training efficiency?

\item \textbf{RQ2}: Can we exploit the coarse-to-fine structure of audio tokens~\cite{kumar2023dac} to further improve the training efficiency of SSM-based TTM?

\item \textbf{RQ3}: Given the potential training efficiency of SSMs, can we further scale down the size of our SSM-based LM to fewer than 100M parameters? 

\end{itemize}

Besides validating the effectiveness of SSMs for TTM, this work also represents one of the first attempts to reduce the cost of TTM training, while fully meeting all the essential openness criteria of the MusGO leaderboard~\cite{batlle2025musgo}. 
Audio samples and the source code are available on our project page.

\section{Background}

\subsection{Text-to-Music (TTM) Generation}
\label{bg:TTM}

Text-to-music (TTM) generation is a conditional music generation task that aims to generate an audio signal 
given a user-provided text prompt, which is a sequence of words. 
Existing approaches largely fall into two paradigms.
The first paradigm relies on pretrained audio codecs~\cite{defossez2022encodec} to convert musical audio into a sequence of discrete \emph{audio tokens} (see Section~\ref{sec_method_overall} for details), enabling text-conditioned sequence modeling using an LM such as the Transformer~\cite{copet2023simple,agostinelli2023musiclm,rouard2024musicgenstyle}. 
The second paradigm
usually employs latent diffusion or flow matching to model continuous embeddings of audio~\cite{evans2025stable,liu2023audioldm,melechovsky2023mustango,prajwal24icml_musicflow,evans2024fast} instead. 
We focus on the first paradigm because we take MusicGen-small as the reference model.
Moreover, we do not consider ``song generation'' models that involve lyrics and vocal generation~\cite{yuan2025yue,ning2025diffrhythm} in this paper.


Existing TTM work tends to focus  on  generation quality and inference speed, rather than  training cost. 
For better generation quality, scaling \emph{up} the model size (e.g., MusicGen-Large has 3.3B parameters~\cite{copet2023simple}) and the training data size is usually found to be helpful. However, the vast computational requirements pose a barrier to democratizing TTM research, restricting from-scratch training of such models to institutions with ample resources.
Moreover, as advocated in MusGO~\cite{batlle2025musgo}, the openness of most SOTA models is hindered by the use of private training data, calling for a more training-affordable and accessible TTM framework.
We respond to this call and explore scaling \emph{down} the model and data sizes in this paper.

\subsection{Efficiency Considerations}
The performance of Transformer architectures is known to benefit from scale, both in terms of data and compute. Classical scaling law studies~\cite{kaplan2020scaling, hoffmann2022training} demonstrate that SOTA performance is often achieved by training increasingly large models on massive datasets, a trend that has also been observed in vision models such as Vision Transformers (ViT)~\cite{touvron2021training, liu2021efficient, cao2022training}. While effective, this paradigm makes modern generative models notoriously data- and compute-hungry.
In parallel, music generative modeling has witnessed rapid progress, with recent studies demonstrating high-quality generation results \cite{evans2025stable, copet2023simple}. As a consequence, efficiency rather than raw modeling capacity has emerged as a critical bottleneck, especially for researchers operating under limited computational budgets or relying on open-source resources.

Efficiency has been discussed from three perspectives in the literature: computational, data, and training efficiency. \emph{Computational efficiency} primarily concerns reducing inference-time cost, the focus of most efficiency-related studies in music generation \cite{schneider2023mo, lam2023efficient, novack2024ditto}. \emph{Data efficiency} aims to reduce the amount of training data required to achieve strong performance \cite{touvron2021training, liu2021efficient, cao2022training}. \emph{Training efficiency} focuses on reducing the cost of optimization during training, such as lowering the number of training iterations, accelerating convergence, or improving hardware utilization \cite{cueto2024framework, yan2024enhancing}. Despite its importance, training efficiency has received comparatively less attention in music generation among the three, where models often require prolonged training schedules and substantial computational resources. 
In light of this gap, it is particularly important to examine training efficiency under realistic and open-resource training settings.

In this work, we consider a practical and open-resource setting, where a text-to-music (TTM) generation model is trained from scratch using publicly available datasets and limited computational resources. Under this setting, the availability of training data is largely fixed, making data efficiency a secondary concern. Instead, training efficiency emerges as the primary bottleneck, as it directly affects the affordability, reproducibility, and accessibility of TTM research. Accordingly, we focus on examining \emph{training efficiency} under a constrained training budget, where efficiency is operationalized by limiting the total number of training iterations (or equivalently, the total number of training tokens). This allows us to evaluate how effectively TTM models can be trained within a fixed training budget, without introducing additional data requirements or computational overhead, while maintaining generation quality.

\begin{figure*}
\centering
\includegraphics[width=1.8\columnwidth]{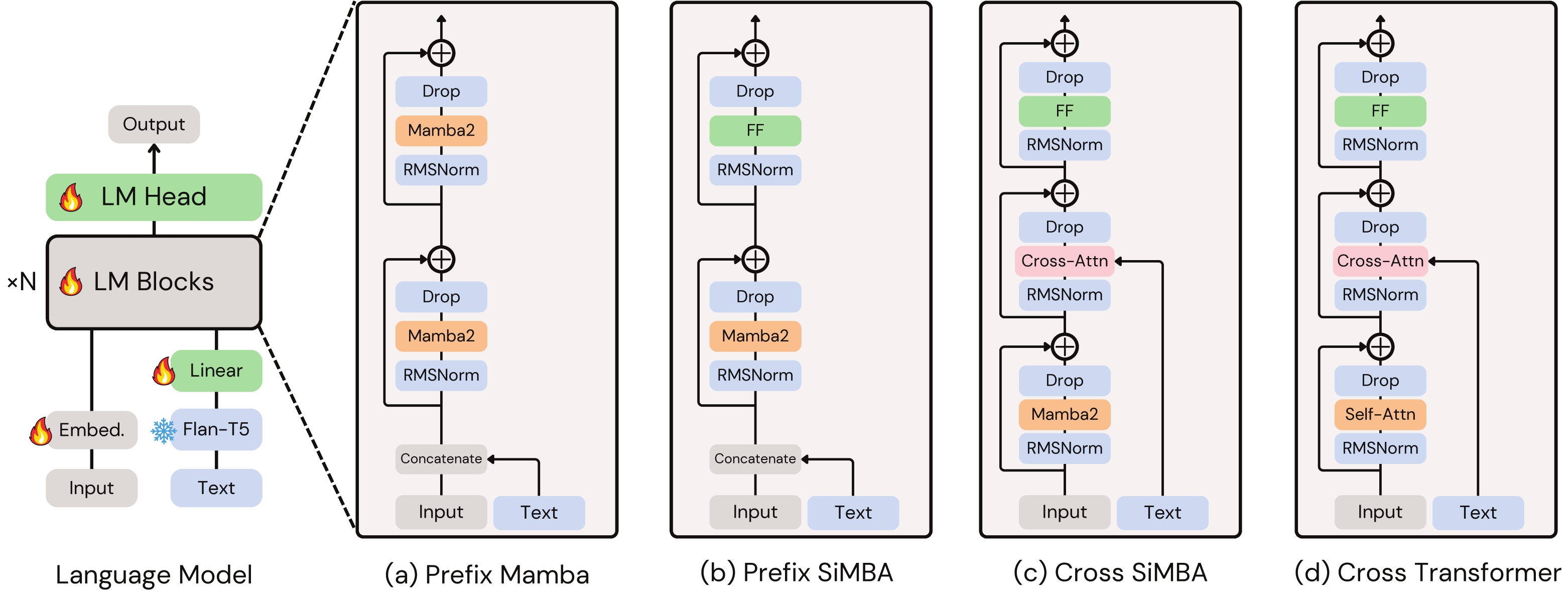}
\caption{
\textbf{LM architectures.} (Left): overall architecture, using the DAC codec~\cite{kumar2023dac} for the input RVQ audio tokens and Flan-T5~\cite{chung2024t5} for processing the input text. (Right)~(a)--(d): the LM variants evaluated, including Mamba-2~\cite{dao2024mamba2}, SiMBA~\cite{patro2024SiMBA}, and Transformer, using either prefix or cross-attention to condition the generation on the input text.}
\label{fig:LM}
\end{figure*}


\subsection{State-Space Models (SSMs)} 
\label{bg:SSM}

SSMs process sequential inputs by compressing the information into a fixed-size hidden state, which is updated incrementally as new tokens arrive. 
This design enables efficient long-range sequence modeling~\cite{gu2025tradeoffs}. 
The underlying mechanism of SSMs can be described by a set of first-order differential equations mapping \emph{continuous} inputs $x(t) \in \mathbb{R}^{p}$ to outputs $y(t) \in \mathbb{R}^{q}$ through a latent state $h \in \mathbb{R}^{n}$:
\begin{equation}
\begin{split}
    h'(t) &= {\rm \textbf{A}} h(t)+{\rm \textbf{B}}x(t) \,,\\
    y(t) &= {\rm \textbf{C}}h(t) \,,
\end{split}
\label{eqa:ssm_1}
\end{equation}
where $\textbf{A} \in \mathbb{R}^{n \times n}$, $\textbf{B} \in \mathbb{R}^{n \times p}$, and $\textbf{C} \in \mathbb{R}^{q \times n}$ are the state, input, and output matrices, respectively.
The same methodology can be applied to \emph{discrete} sequences of tokens 
with some techniques to obtain discretized matrices $\overline{\textbf{A}}$ and $\overline{\textbf{B}}$ \cite{gu2021lssl,gu2021s4}.
This leads to the following recurrent discrete SSM mapping $\mathbf{x}_t \in \mathbb{R}^{T}$ (which can be viewed as an embedding representation of $x_t$) to $\mathbf{y}_t \in \mathbb{R}^{T}$ with $\mathbf{h}_t \in \mathbb{R}^{T \times D}$, making the inference as fast as RNNs.
\begin{equation}
\begin{split}
    \mathbf{h}_t &= {\overline{\textbf{A}}} \mathbf{h}_t+{\overline{\textbf{B}}}\mathbf{x}_t \,,\\
    \mathbf{y}_t &= {\textbf{C}}\mathbf{h}_t \,.
\end{split}
\label{eqa:ssm_2}
\end{equation}
Besides, its linear time-invariant (LTI) 
property enables pre-computation of a 
conv 
kernel 
$\overline{\textbf{K}}=(\textbf{C}\overline{\textbf{B}}, \textbf{C}\overline{\textbf{AB}}, \textbf{C}\overline{\textbf{A}}^2\overline{\textbf{B}}, \dots)$,
facilitating efficient parallel training like CNNs.

Newer \emph{selective SSMs} such as Mamba~\cite{gu2023mamba}, Mamba-2~\cite{dao2024mamba2} introduce a selective scan mechanism, where the matrices $\overline{\textbf{B}}_t$, $\textbf{C}_t$, and step size $\Delta_t$ depend dynamically on the input $\mathbf{x}_t$, improving modeling ability and making the overall model dynamic linear RNNs.
To improve training efficiency,
Mamba-2~\cite{dao2024mamba2} 
introduces a structured state space duality layer, 
computing $\overline{\textbf{A}}$, $\overline{\textbf{B}}_t$, and $\textbf{C}_t$ in parallel via a single linear layer.

By integrating linear updates, parallel scans, and the selectivity mechanism, Mamba and its variants~\cite{gu2023mamba,dao2024mamba2} have emerged as strong alternatives to the Transformers for building LMs,
in language~\cite{lieber2024jamba, ren2025samba}, vision~\cite{liu2024vmamba, ma2024tinyvim}, and other domains. 
Their application to the music domain is more recent and has been limited to the generation of symbolic MIDI music~\cite{chen2025musicmamba,zhang2025mambadiff} and short audio sound effects~\cite{colombo2025mambafoley}, often under a simpler setting with smaller datasets (e.g.,  6 hours of audio~\cite{colombo2025mambafoley}), and without a deeper investigation of different model variants. In particular, to our best knowledge, little work has employed SSMs for the more challenging task of TTM.



\section{Proposed Ideas}

\subsection{SSM-based LM for Text-to-music Generation}


To explore training-efficient architectures in response to \textbf{RQ1}, we propose three SSM-based LMs for TTM leveraging Mamba-2 \cite{dao2024mamba2} and SiMBA \cite{patro2024SiMBA}, and  empirically compare their performance with that of a Transformer baseline (using FlashAttention \cite{dao2022flashattention} and the same number of $N$ LM blocks) in our experiments. We set $N=24$ by default so that the Transformer baseline matches MusicGen-small with a total 24 layers. (We further investigate in \textbf{RQ3} whether $N$ can be reduced to 6 for SSMs.) 
As depicted in Figure~\ref{fig:LM}, the three SSM variants employ different non-linear components (detailed below), and two different text-conditioning schemes:
1) the \emph{Prefix} approach that simply prepends the text prompt before the sequence of audio tokens, and 2) the \emph{Cross}-attention approach that employs cross-attention layers following the common approach used in the Transformers. 

\begin{itemize}
\item \textbf{Prefix\,Mamba}: As shown in Figure~\ref{fig:LM}(a), Prefix\,Mamba is a purely sequence-mixing SSM-based language model that exclusively employs Mamba-2 layers~\cite{dao2024mamba2} as the core modeling component in each LM block. It adopts the prefix-tuning strategy to incorporate text conditions.
Notably, this architecture contains \emph{no} channel-mixing transformations (e.g., feedforward/MLPs), relying solely on the sequence-mixing dynamics of the Mamba-2 blocks. This design enables a direct investigation into the expressiveness and efficiency of purely sequence-mixing SSMs for TTM.


\item \textbf{Prefix\,SiMBA} introduces 
non-linearity by replacing one of the two Mamba-2 layers per LM block of Prefix\,Mamba with a feedforward block for ``channel mixing,'' as shown in Figure~\ref{fig:LM}(b). 
We name this variant SiMBA because the idea of channel mixing originates from the SiMBA model developed for the vision domain~\cite{patro2024SiMBA}. However, 
unlike the original paper, which uses the more sophisticated Einstein FFT (EinFFT) for channel mixing, we opt for a simpler structure using the basic feedforward layers here.
This design allows us to isolate and examine the contribution of non-linearity in SSMs.

\item \textbf{Cross\,SiMBA} replaces the prefix-based conditioning of Prefix\,SiMBA with an additional multi-head cross-attention layer to incorporate text information, as shown in Figure~\ref{fig:LM}(c). 
Compared to Prefix\,SiMBA, where the text condition is fused into the sequence through prefix tokens and becomes \emph{part of the SSM state}, Cross\,SiMBA \emph{decouples} the text representation from the hidden state of the SSM and integrates it via cross-attention. We propose this design in order to study whether attention-based fusion is more effective than directly updating the SSM state at every step when dealing with cross-modal feature integration in an SSM-based LM for TTM.

\item \textbf{Cross\,Transformer}: As shown in Figure~\ref{fig:LM}(d), Cross\,Transformer serves as a baseline model that adopts the classic architecture of each LM block consisting of multi-head self-attention, multi-head cross-attention, and feedforward layers.
As a widely adopted baseline, it enables a fair efficiency-performance comparison with other variants.
Moreover, as Cross\,SiMBA (Figure \ref{fig:LM}(c)) and Cross\,Transformer (Figure \ref{fig:LM}(d)) share much in common, comparing these two LMs directly contrasts the state-based recurrence of Mamba-2 with the self-attention mechanism of the Transformer for sequence mixing.
\end{itemize}

\subsection{Two-Stage SSM/Diffusion Hybrid Design}
\label{sec_method_overall}

Unlike text tokens, audio tokens typically require \emph{multiple} discrete codes to represent each audio segment faithfully.
Specifically, 
neural audio codecs such as EnCodec~\cite{defossez2022encodec} and DAC~\cite{kumar2023dac} 
employ residual vector quantization (RVQ), 
which recursively computes the residual after quantization for a number of $K$ times, using $K$ quantization codebooks $\{Q_1,\dots, Q_{K}\}$.
Accordingly, each chunk of an audio signal is represented by $K$ audio tokens: $\{x_t^1,\dots,x_t^{K}\}$, where each $x_t^k$ serves as an index into the $k$-th  codebook $Q_k$, and $t$ denotes time.
The token representation of the entire music signal is therefore 
$\{x_1^1,x_1^2,\dots,x_1^{K},x_2^1,\dots\}$.

The token sequence for music can be fairly long, posing a challeng for language modeling.
For instance, for audio fidelity, the DAC~\cite{kumar2023dac} uses ${K}=9$ codebooks and operates at a frame rate of about 86 chunks per second.
To reduce the sequence length $n$ for better training and inference efficiency, MusicGen~\cite{copet2023simple} proposes to generate the $K$ quantization layers of tokens for the same chunk at once, or using a ``delay'' pattern, reducing the sequence length by a factor of $K$.

However, despite these optimizations, modeling tokens from the $K$ hierarchical codebooks concurrently remains a significant challenge for language models. A recent study~\cite{wang2025language} identifies orthogonal properties among tokens from different codebook layers, which complicates the training process and impedes convergence.

In contrast to single-stage generation, this work explores an alternative modeling paradigm. By design, Residual Vector Quantization (RVQ) yields a structured hierarchy of quantized tokens, ranging from \emph{coarse} to \emph{fine}. Recent studies~\cite{lanzendorfer2025coarse} leverage this inherent property to develop multi-stage generation frameworks.

The first quantization layer of tokens, $x^{(1)}:=\{x_1^1,x_2^1,\dots,x_t^1,\dots\}$, represents the coarsest information of the audio signal and carry much of the so-called semantic information that corresponds to the text prompt. 
The remaining (finer-grained) quantization layers, $x^{(2:K)}:=\{x_1^{2},x_1^{3},\dots,x_1^K,x_2^{2},x_2^{3},\dots\}$, are mainly responsible for enhanced audio fidelity.
Therefore, instead of modeling all the $K$ quantization layers with a single model, it might be more efficient to decompose the generation into two stages, with the first-stage model \emph{generating only the coarsest quantization layer}, and the second ``separate'' model generating the tokens of the remaining layers. 

The codec-based two-stage generation paradigm was previously explored in DiscoDiff~\cite{lanzendorfer2025coarse}, which employs non-autoregressive (NAR) latent diffusion models (LDMs) for both stages.
We adapt the idea and propose to use an autoregressive (AR) SSM-based LM for the first stage instead.
The second stage LDM was pre-trained on the same dataset as the first stage, strictly adhering to the training procedures outlined in the official DiscoDiff implementation. This decision aligns with our commitment to open science and experimental reproducibility; by utilizing a fully public and accessible dataset, we ensure that our hybrid framework is evaluated in a transparent and standardized environment.
This leads to a two-stage \textbf{SSM/diffusion} hybrid design for TTM.

Formally, the first-stage model $f(\cdot)$, parametrized by $\theta$, considers the input text $c$ and generates a rudimentary audio sequence successively in chronological order:
\begin{equation}
    p(x_t^{1} | x_{<t}^{1}, \, c) = f(x^{(1)}, c; \theta) \,,
    \label{eq:first_stage}
\end{equation}
where $x_{<t}^{1}$ represents the sequence of first-layer tokens for all the chunks prior to the current $t$-th chunk. 
Given $c$ and the entire rudimentary audio across all the chunks, the second-stage model $g(\cdot)$, parametrized by $\vartheta$, aims to enhance the audio quality by predicting the  finer-grained tokens:
\begin{equation}
    p(x^{(2:K)} | x^{(1)}, \, c) = g({x}^{(1)}, c; \vartheta) \,.
    \label{eq:second_stage}
\end{equation}
The second-stage model is separately pre-trained and can be used regardless of how the first-stage model is constructed.

\textbf{RQ2} empirically investigates whether such a two-stage design improves training efficiency by comparing SSM/diffusion hybrid with a single-stage SSM that models all $K$ tokens concurrently, and with a Transformer/diffusion two-stage variant that uses the Transformer as the first-stage model.

Interestingly, TinyViM~\cite{ma2024tinyvim} suggests that Mamba blocks predominantly capture low-frequency components in the input sequence in the vision domain. Similarly, we observe in our pilot study and unofficial listening that the first DAC layer indeed captures semantic-related low-frequency information of music, adding support to the use of SSMs for TTM.

\subsection{Curating a Public Dataset for TTM Training}
Our training data is derived from Jamendo~\cite{bogdanov2019mtg}, a public collection of 54,894 tracks with Creative Commons (CC) licenses. 
We use 44.1kHz sampling rate and mono audio, 
splitting the tracks into non-overlapping 30s clips and removing the vocals using HTDemucs~\cite{rouard2023hybrid}.
LP-MusicCaps~\cite{doh2023lp} generates three 10s captions per clip, which are rephrased by LLaMA-3~\cite{grattafiori2024llama} into a single description, while excluding clips with residual vocal cues. We discard clips with more than 15s of silence to ensure musical richness. 
To train our LMs, we randomly select one clip per track, resulting a training collection of 457.45 hours in total.
This dataset is small compared to the private data used by the official MusicGen-small (with 20K hours), but it is fully CC-licensed and can be shared publicly.

\section{Experimental Setup}

\subsection{Training and Inference Setup}
We choose DAC~\cite{kumar2023dac} ($K=9$) as the audio codec in our implementation. 
To encode text information of the input prompt, we use a pretrained Flan-T5 Base model~\cite{chung2024t5}.

We train all our LMs from scratch on a single RTX 3090 GPU for up to 10k training steps, as mentioned in the introduction. 
We implement the four LMs introduced in Figure~\ref{fig:LM} on our own, all using $N=24$ LM blocks with a hidden dimension of 1,024, a Mamba-2 block state dimension of 512 (if present), and 8 heads for the self-and cross-attention layers (if present), with a dropout rate of 0.3, a batch size of 4, and 32 gradient accumulation steps. 
The feedforward (FF) layer in the SiMBA variants (i.e., \ref{fig:LM}(b) and \ref{fig:LM}(c)) consists of two linear transformations 
and a GELU non-linear activation and dropout regularization in between.
We use the AdamW optimizer~\cite{loshchilov2017adamW} with learning rate 1e--4 and weight decay 2e--2, betas (0.9, 0.999), a cosine annealing scheduler and a 100-step warm-up.


At inference, all generation models are configured with a classifier-free guidance scale of 3. For LMs, we employ top-k sampling retaining the top 250 tokens and set temperature $\tau =$ 1.2. For the LDM, we use DDPM~\cite{ho2020ddpm} for denoising with the denoising step $m= 100$.

\subsection{Evaluation Protocol}

For the evaluation dataset, we use MusicCaps~\cite{agostinelli2023musiclm}, retrieving 5,367 out of the 5,521 text-audio pairs from the internet (the rest are not accessible), ensuring no overlaps between MusicCaps and our training data.
The audio tracks are similarly sampled at 44.1kHz and converted to mono.
As our training data is vocal-free, we exclude captions with vocal-related terms and randomly sample 200 and 50 clips respectively for our experiment on 10s and 25s generation tasks.
We choose MusicCaps over alteratives such as the Song Describer Dataset~\cite{manco2023sdd}, because its 10s text-audio pairs provide 
richer musical details in text. 

For objective evaluation, we use FD$_{openl3}$, KL, CLAP for audio quality/realism, semantic similarity, and the alignment of text-audio pairs, respectively.

\begin{itemize}
\item \textbf{Fréchet Distance (FD$_{openl3}$)} \cite{ma2024look} evaluates the quality and realism of the generated audio by computing the Fréchet distance between $L^3$-Net embeddings of the generated samples and those from a reference set (i.e., MusicCaps). 
Lower FD indicates that the generated audio is closer to real audio in the reference set in distributions, reflecting higher perceptual quality and realism.

\item \textbf{Kullback-Leibler divergence (KL)} quantifies the semantic similarity between generated and reference audio by comparing their music tag distributions predicted by PaSST, a SOTA music tagging model~\cite{koutini2021efficient}. Following~\cite{evans2024fast},
long-form audio is segmented, and KL scores are averaged across segments. Lower KL indicates higher semantic consistency.

\item \textbf{CLAP}~\cite{elizalde2023clap} measures cross-modal alignment by computing the cosine similarity between audio and text embeddings obtained from the CLAP model trained on LAION’s music-caption dataset. Higher CLAP indicates stronger text-audio semantic relevance.

\end{itemize}

We further perform two subjective listening tests (one for RQ2 and the other for RQ3), with informed consent obtained to evaluate the perceptual quality of the generated audio. 
Each session of the listening test has three questions, each presenting a unique text prompt sampled from MusicCaps, along with anonymized, randomly ordered audio clips generated by different models.
Participants rate each clip on a 5-point Likert scale (1 = poor, 5 = excellent) across the following dimensions:
\begin{itemize}
\item \textbf{Consistency}: Measures local and global musical coherence, including melody, rhythm, and overall structure.
\item \textbf{Alignment}: Assesses how well the input text prompt is reflected in the generated audio, for descriptions related to either explicit attributes (e.g., instrumentation, tempo) or abstract semantics (e.g., mood, style).
\item \textbf{Overall}: Evaluates the musicality, naturalness, and general listenability of the generated audio.
\end{itemize}

\begin{table}
\caption{Objective evaluation result of the four LMs depicted in Figure~\ref{fig:LM} for generating 10s clips at an early, 20k-th training step, using only one layers of DAC tokens; $\downarrow/\uparrow$ indicates lower/higher the better, and best results highlighted.
}
\setlength{\tabcolsep}{1mm}
\centering
\begin{adjustbox}{width=0.9\columnwidth}
\begin{tabular}{l|c|rrr}
\toprule 
\multirow{2}{*}{\textbf{Model}} & \textbf{Model} & \multicolumn{3}{c}{10s-long clips}\\
~ &\textbf{size}&\textbf{FD$_{openl3}$ ${\downarrow}$}&\textbf{KL ${\downarrow}$}&\textbf{CLAP ${\uparrow}$}\\
\midrule
GT (1 layer)& ---
&59.3433&1.4950&0.1950\\
\midrule
Prefix\,Mamba&357 M
&107.0340&1.8837&\underline{0.1450}\\
Prefix\,SiMBA&281 M
&\underline{102.7248}&\underline{1.5106}&\textbf{0.1776}\\
Cross\,SiMBA&381 M
&118.4306&\textbf{1.3012}&0.1240\\
Cross\,Transformer&306 M
&\textbf{99.2973}&3.1077&0.1150\\
\bottomrule
\end{tabular}
\end{adjustbox}

\label{table:exp1-1}
\end{table}

\begin{figure}
  \centering
  \includegraphics[alt={LM performance along the training steps},width=0.9\columnwidth]
  {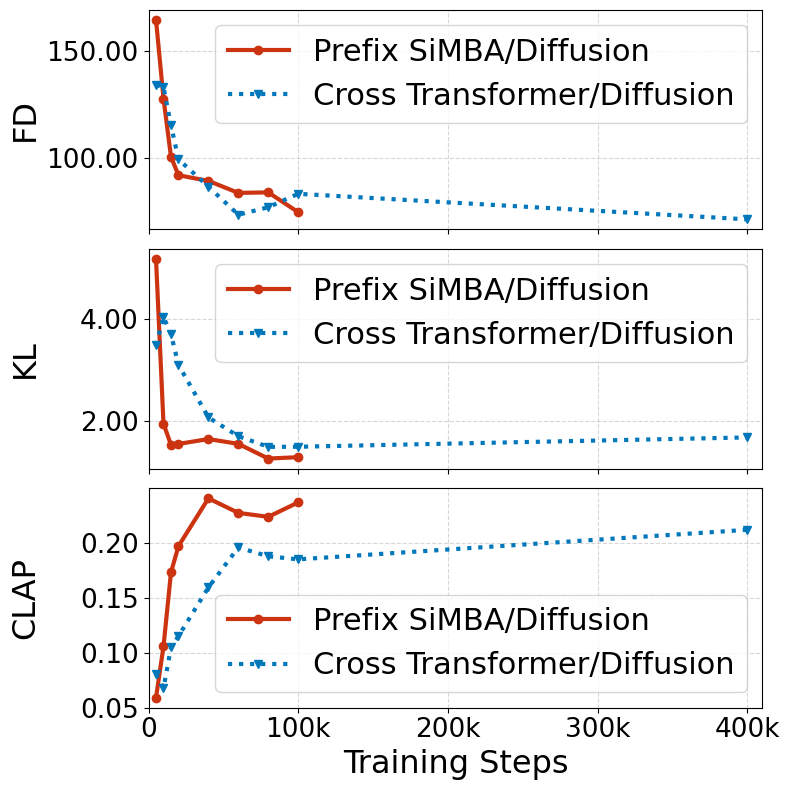}
  \caption{Objective result of two selected LMs as a function of training steps. Performance is evaluated across training steps within the two-stage framework (Diffusion stage is frozen). Prefix\,SiMBA demonstrates superior training efficiency.}
  \label{fig:score_training_steps}
\end{figure}

\begin{figure}[t]
    \centering
    \begin{adjustbox} {width=0.9\columnwidth}
    \subfloat[SiMBA/Diffusion (100k)]{
        \includegraphics[width=0.5\linewidth]{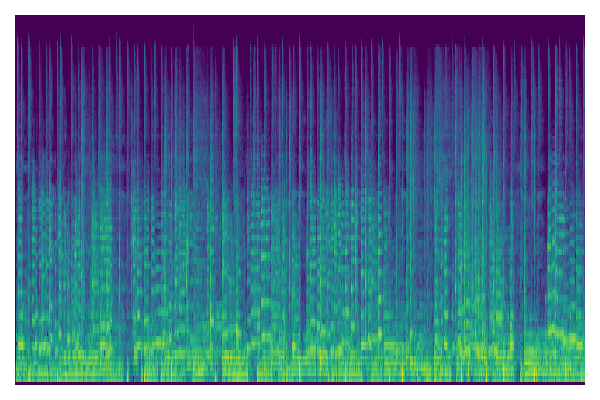}
    }\hfill
    \subfloat[SiMBA/Diffusion (50k)]{
        \includegraphics[width=0.5\linewidth]{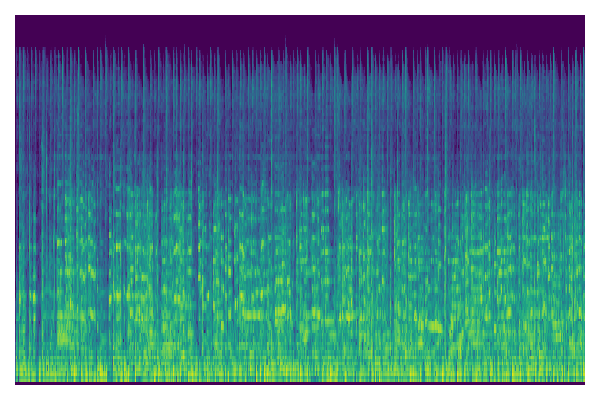}
    }
    \end{adjustbox}
    \begin{adjustbox} {width=0.9\columnwidth}
    \subfloat[Transformer/Diffusion (400k)]{
        \includegraphics[width=0.5\linewidth]{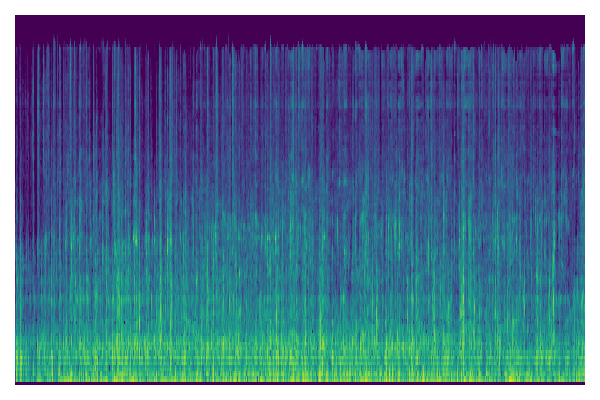}
    }\hfill
    \subfloat[Transformer/Diffusion (100k)]{
        \includegraphics[width=0.5\linewidth]{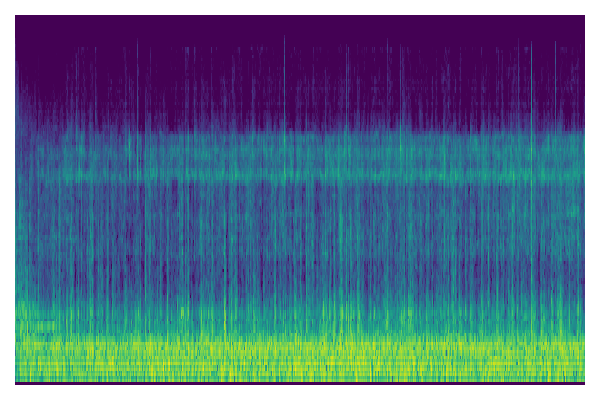}
    }
    \end{adjustbox}
    \caption{Comparison of generated Mel-spectrograms from SiMBA-based and Transformer-based diffusion models using the same text prompt describing a jazz-reggae concert with multiple instruments. Sub-figures (a)-(d) illustrate the visual consistency and spectral detail across different architectures and training iterations.}
    \label{fig:qualitative}
\end{figure}

\begin{table*}
\caption{Objective evaluation results comparing our two-stage model `Prefix\,SiMBA/Diffusion' with its single-stage variant (`Prefix\,SiMBA$_9$'), its two-stage Transformer counterpart (`Transformer/Diffusion'), and the official MusicGen-small, for full-band audio generation (using all the $K$ quantization layers) of either 10s or 25s long. MusicGen-small is trained by its authors using a private data of 20K hours, while ours are trained using the 0.45K hours Jamendo data. The table also reports the number of parameters for each stage; the shorthands `T' and `F' denote trainable and frozen modules, respectively.}
\centering
\begin{adjustbox} {width=1.8\columnwidth}
\begin{tabular}{l|ccc|rrr|rrr}
\toprule 
\multirow{2}{*}{\textbf{Model}} & \multirow{2}{*}{\textbf{Data}} & \multicolumn{2}{c}{\textbf{\# of Parameters}} & \multicolumn{3}{c}{10s-long clips}& \multicolumn{3}{c}{25s-long clips} \\ 
~ & ~& \textbf{Stage~1}& \textbf{Stage~2} &\textbf{FD$_{openl3}$ ${\downarrow}$}&\textbf{KL ${\downarrow}$}&\textbf{CLAP ${\uparrow}$}&\textbf{FD$_{openl3}$ ${\downarrow}$}&\textbf{KL ${\downarrow}$}&\textbf{CLAP ${\uparrow}$}\\
\midrule

Official MusicGen-small & Private
&300M (T)&--- 
&109.1424&1.8371&\textbf{0.2422}
&125.8558&1.3879&0.2234\\
\midrule
{Prefix\,SiMBA$_9$}& Jamendo 
&300M (T)&---
&98.6492&\textbf{1.1952}&0.2360
&107.2963&\textbf{0.6902}&\underline{0.2280}\\
Transformer/Diffusion& Jamendo 
&306M (T)&1.2B (F)
&\underline{83.1081}&1.4994&0.1852
&\underline{97.3981}&1.2494&0.1630\\
Prefix\,SiMBA/Diffusion& Jamendo 
&281M (T)&1.2B (F)
&\textbf{74.4559}&\underline{1.2948}&\underline{0.2370}
&\textbf{85.2023}&\underline{0.9408}&\textbf{0.2495}\\
\bottomrule
\end{tabular}
\end{adjustbox}

\label{table:exp_2}
\end{table*}

\section{Experimental Results}
\label{section:Experimental Results}

We present a series of experiments designed to answer the three RQs mentioned in the introduction section.

\subsection{SSMs versus Transformers}
\label{sec:exp:1}

We first address \textbf{RQ1} and evaluate only the first-stage model $f(\cdot)$ by comparing the four LMs in Figure~\ref{fig:LM}, generating the first-layer DAC tokens. Since DAC can reconstruct audio from arbitrary numbers of quantization layers, the second-stage model $g(\cdot)$ is not required, allowing us to isolate and assess the modeling capability of the LM. 
We consider a `10s audio generation' task here and generate audio of 10-seconds long.
Because we only use one quantization layer here, the audio quality of the generated audio would be inherently limited. Therefore, we include the result of real audio reconstructed using only the first-layer DAC tokens for comparison. 
Note that in this experiment, all the LMs (including the three SSM variants and the Transformer baseline) are trained by ourselves under the same training configuration and using our own training data.
Namely, the Transformer baseline here is not the official MusicGen-small.

Table~\ref{table:exp1-1} reports the objective results of the 10s audio generation task for the four LMs trained for 20k training steps, alongside the oracle result of the reconstructed audio  (GT). 
We select the checkpoints at 20k training steps in this evaluation to assess the training efficiency of the models.

Table~\ref{table:exp1-1} leads to the following observations.
First, comparing Prefix\,Mamba (corresponding to the architecture in Figure~\ref{fig:LM}(a) and Prefix\,SiMBA (Figure~\ref{fig:LM}(b)) highlights the critical role of \emph{channel-mixing layers} in the fusion process. By introducing MLP-based mixing on top of state and condition, Prefix\,SiMBA not only achieves improved performance across all three objective metrics but also enables a more compact model size, showcasing its effectiveness in enhancing both efficiency and expressiveness.

Second, a comparison between Prefix\,SiMBA (Figure~\ref{fig:LM}(b)) and Cross\,SiMBA (Figure~\ref{fig:LM}(c)) shows that these two models perform similarly for KL, but Prefix\,SiMBA achieves better FD and CLAP score.  
This suggests that incorporating the text condition \emph{inside} the SSM state is more effective for cross-modal text-audio alignment, than incorporating the text condition \emph{outside} the SSM state via cross attention. 
With the same number of $N=24$ LM blocks, Prefix\,SiMBA also has a much smaller model size than Cross\,SiMBA.
This demonstrates the effectiveness of SSMs in sequence modeling.

Third, comparing Cross SiMBA (Figure~\ref{fig:LM}(c))  to Cross Transformer (Figure~\ref{fig:LM}(d)), we note that the latter performs worse in terms of KL and CLAP, actually the worst among the four LMs. 
Under the same training budget and data, the matched Transformer architecture converges more slowly than the proposed SSM-based models.

Altogether, the observations from Table~\ref{table:exp1-1} validate the efficacy of incorporating lightweight non-linearities (Prefix SiMBA) and suggest limitations of both purely linear models (Prefix Mamba) and naive modality fusion (Cross\,SiMBA), answering the \textbf{RQ1} listed in the introduction.

Figure~\ref{fig:score_training_steps} further plots the objective scores as a function of the training steps of LMs, under the same two-stage paradigm.
Prefix\,SiMBA exhibits substantially faster convergence and better early-stage performance than a matched Transformer baseline under the same constrained training setup, demonstrating superior training efficiency.
Notably, different metrics exhibit distinct convergence behaviors. The KL for Prefix\,SiMBA reaches a plateau as early as at the 15k step, suggesting that the model rapidly captures the overall statistical structure of the audio tokens. In contrast, the FD continues to decrease beyond 100k, indicating that fine-grained audio quality and diversity may still be improving and have not yet fully converged. The CLAP similarity, which reflects semantic alignment between text and audio, peaks and stabilizes around 40k steps, highlighting the model’s early development of strong cross-modal understanding.
In practice, we observe that under a constrained training budget of 100k steps, the Transformer baseline fails to consistently generate perceptually plausible audio. To ensure a fair comparison, the Transformer baseline is therefore trained for up to 400k steps until it reliably produces coherent and audible music. Even under this extended training regime, its performance remains comparable to that of Prefix,SiMBA trained with 100k steps or fewer.
These results suggest that Prefix\,SiMBA learns alignment and distributional structure early in training, while continued improvement in perceptual fidelity (as measured by FD) may require longer optimization.



\begin{table}
\caption{Subjective evaluation results (mean~$\pm$~standard deviation) comparing the official MusicGen-small with the Prefix\,SiMBA/Diffusion (denoted as `SSM' here) and Prefix\,Transformer/Diffusion (`Transformer'), trained with different training step budgets (50k, 100k, and 400k steps), for generating 25s audio clips.}
\setlength{\tabcolsep}{1mm}
\centering
\begin{adjustbox}{width=0.9\columnwidth}
\begin{tabular}{l|ccc}
\toprule
\textbf{Model}&\multicolumn{1}{c}{\textbf{Consistency}}&\multicolumn{1}{c}{\textbf{Alignment}}&\multicolumn{1}{c}{\textbf{Overall}}\\
\midrule

MusicGen-small & \textbf{3.73\footnotesize{~$\pm$1.06}}&\underline{3.65\footnotesize{~$\pm${1.05}}}&\textbf{3.60\footnotesize{~$\pm${1.04}}}\\
\midrule
SSM (100k) &\underline{3.37\footnotesize{~$\pm${1.05}}}&\textbf{3.65\footnotesize{~$\pm${0.89}}}&\underline{3.42\footnotesize{~$\pm${1.00}}}\\
SSM (50k) &{2.98}\footnotesize{~$\pm${0.99}}&{2.85}\footnotesize{~$\pm${1.06}}&{2.77}\footnotesize{~$\pm${1.01}}\\
Transformer (400k) &{2.92}\footnotesize{~$\pm${0.92}}&{3.03}\footnotesize{~$\pm${1.08}}&{2.88}\footnotesize{~$\pm${1.10}}\\
Transformer (100k) &{2.40}\footnotesize{~$\pm${1.10}}&{2.38}\footnotesize{~$\pm${1.08}}&{2.27}\footnotesize{~$\pm${1.17}}\\

\bottomrule
\end{tabular}
\end{adjustbox}

\label{table:sub_1}
\end{table}

\begin{table*}
\caption{Objective and subjective results when scaling down the  size of the proposed SSM-based  model, for 25s generation.}
\centering
\begin{adjustbox}{width=1.8\columnwidth}
\begin{tabular}{l|rrr|ccc}
\toprule 
\textbf{Model}&\textbf{FD$_{openl3}$ ${\downarrow}$}&\textbf{KL ${\downarrow}$}&\textbf{CLAP ${\uparrow}$}
&\multicolumn{1}{c}{\textbf{Consistency}}&\multicolumn{1}{c}{\textbf{Alignment}}&\multicolumn{1}{c}{\textbf{Overall}}\\
\midrule
Prefix\,SiMBA/Diffusion (24 LM blocks) & 85.2023 & 0.9408 & 0.2495
&\textbf{3.67\footnotesize{~$\pm${1.01}}}&\textbf{3.64\footnotesize{~$\pm${1.08}}}&\textbf{3.61\footnotesize{~$\pm${0.99}}}\\
Prefix\,SiMBA/Diffusion (12 LM blocks) & 88.1027 & 1.1073 & 0.2547
&\underline{{3.24}\footnotesize{~$\pm${1.01}}}&\underline{{3.25}\footnotesize{~$\pm${1.04}}}&\underline{{3.18}\footnotesize{~$\pm${1.02}}}\\
Prefix\,SiMBA/Diffusion (6 LM blocks) & 82.0459 & 0.9284 & 0.2442 
&{2.54}\footnotesize{~$\pm${1.11}}&{2.67}\footnotesize{~$\pm${1.10}}&{2.37}\footnotesize{~$\pm${1.09}}\\
\bottomrule
\end{tabular}
\end{adjustbox}

\label{table:exp3}
\end{table*}

\subsection{Single-stage versus Two-stage Generation}
\label{sec:exp:2}

Next, we address \textbf{RQ2} and investigate whether leveraging the coarse-to-fine two-stage generation framework leads to improved training efficiency for SSM-based TTM, compared to end-to-end single-stage modeling. 
In particular, we aim to generate full-resolution audio with all the $K$ quantization layers.
Moreover, we aim to compare our best model with the official MusicGen-small when generating either 10s audio or 25s audio. The latter  task (`25s audio generation') is included to assess longer-form generation.
In this experiment, we train from scratch the following three models for 100k training steps using our Jamendo dataset.
\begin{itemize}
\item \textbf{Prefix\,SiMBA$_9$}: This model performs LM-based generation over all $K$ layers of DAC token sequences. As in MusicGen~\cite{copet2023simple}, we adopt the \textit{delay} pattern, which right-shifts the $k$-th layer by $k$ steps and sums them to factorize the joint distribution. Compared to the \textit{flattening} pattern (which concatenates all layers into a long $K \times T$ sequence, for an audio with $T$ chunks), the delay approach yields a shorter input length of $K + T - 1$, at the cost of slightly increased modeling complexity. 

\item \textbf{Prefix\,SiMBA/Diffusion}: This two-stage approach follows the design in Section~\ref{sec_method_overall}. It uses Prefix\,SiMBA, the best SSM-based LM as shown in Section~\ref{sec:exp:1}, as the first-stage model $f(\cdot)$ to generate the first quantization layer of DAC tokens, and a pre-trained LDM as the second-stage model $g(\cdot)$ to infer the finer-grained tokens.

\item \textbf{Cross\,Transformer/Diffusion}: This model uses the same two-stage framework as Prefix\,SiMBA/Diffusion, but replaces Prefix\,SiMBA with a Cross\,Transformer (cf. Figure~\ref{fig:LM}(d)) as the first-stage model $f(\cdot)$.
\end{itemize}
Despite a considerably smaller training set than MusicGen-small and trained for only 100k steps, both Prefix\,SiMBA$_9$ and Prefix\,SiMBA/Diffusion achieve performance competitive with MusicGen-small, validating the efficiency of the proposed models in resource-constrained scenarios. 
When comparing Prefix\,SiMBA$_9$ and Prefix\,SiMBA/Diffusion, we see that the two-stage model achieves significantly better performance in both FAD and CLAP scores. The result supports the effectiveness and efficiency of the proposed coarse-to-fine two-stage generation framework for an SSM-based LLM, answering \textbf{RQ2}. Interestingly, Prefix\,SiMBA$_9$ achieves lower KL divergence, consistent with our earlier observation in 
Section~\ref{sec:exp:1} 
that SSMs introduce strong inductive biases beneficial for certain statistical measures. 
Moreover, within the two-stage models, Prefix\,SiMBA/Diffusion consistently outperforms Cross\,Transformer/Diffusion across all the evaluation metrics, indicating that Prefix\,SiMBA provides a more efficient conditional representation, particularly in resource-constrained settings.
Finally, only Prefix\,SiMBA/Diffusion benefits from longer sequence generation, showing improved CLAP scores when generating 25-second clips compared to 10-second ones. This suggests that Prefix\,SiMBA is especially effective for modeling longer audio.

Table~\ref{table:sub_1} shows the subjective evaluation results averaged from 20 participants comparing the official MusicGen-small with Prefix\,SiMBA/Diffusion and Transformer/Diffusion, where the first-stage language models are trained under different step budgets, including 50k, 100k, and 400k steps.
We see that Prefix\,SiMBA/Diffusion\,(100k) achieves perceptual quality comparable to MusicGen-Small, especially for the Alignment, indicating stronger semantic correspondence between the input text and generated audio. This observation aligns with our CLAP-based objective evaluation, where Prefix\,SiMBA consistently achieves higher cross-modal similarity. 
While MusicGen-Small attains slightly higher scores in Consistency and Overall quality, this may be attributed to its significantly larger training dataset and longer training schedule—underscoring a trade-off between training cost and generation quality.

Table~\ref{table:sub_1} further shows that Prefix\,SiMBA/Diffusion consistently outperforms Transformer/Diffusion across all subjective metrics when trained under comparable step budgets. Notably, even the 50k-step Prefix\,SiMBA/Diffusion model exhibits performance comparable to that of the 400k-step Transformer baseline, underscoring the superior training efficiency of the SiMBA language model.

We further conducted a qualitative comparison between Prefix\,SiMBA/Diffusion and Transformer/Diffusion using mel spectrograms. Both models were conditioned on the same prompt.\footnote{
``This is the recording of a jazz reggae concert. There is a saxophone lead playing a solo. There is a keyboard and an electric guitar playing the main tune with the backing of a bass guitar. In the rhythmic background, there is an acoustic reggae drum beat. The atmosphere is groovy and chill. This piece could be playing in the background at a beach. It could also be included in the soundtrack of a summer/vacation/tropical themed movie.''
}
As shown in Figure~\ref{fig:qualitative}, Prefix\,SiMBA/Diffusion (Figures~\ref{fig:qualitative}(a),\,(b)) exhibits sharper vertical transients, reflecting the percussive characteristics of the drums and electric guitar. It also better preserves high-frequency details and harmonics, crucial for the saxophone timbre, compared to Transformer/Diffusion (Figures~\ref{fig:qualitative}(c),\,(d)).
While Transformer/Diffusion gradually captures some spectral details with extended training, even after 400k steps it remains inferior to Prefix\,SiMBA/Diffusion trained for only 50k steps. We hypothesize that this disparity arises because jazz-reggae prompts are relatively rare in the Jamendo dataset. As a data-hungry model, the Transformer fails to fully capture the characteristics of such uncommon descriptions within the given number of iterations. Consequently, this example highlights the training efficiency advantage of Prefix\,SiMBA/Diffusion for capturing complex musical textures with limited data.

We note that training Prefix\,SiMBA/Diffusion\,(100k) requires approximately $10^{19}$ FLOPs, which is about one order of magnitude smaller than the FLOPs required to train the official MusicGen-small.



\subsection{Scaling Down the SSM-based LM}

To investigate the parameter efficiency of the proposed architecture and address \textbf{RQ3}, we progressively reduce the number of LM blocks (i.e., $N$) in the Prefix\,SiMBA language model and examine how generation quality varies as the model becomes more compact. While reducing model capacity generally leads to performance degradation, this experiment aims to assess whether the strong sequence modeling capability of SSMs allows Prefix,SiMBA to preserve generation quality with substantially fewer trainable parameters.
Specifically, in addition to the default configuration with 24 LM blocks ($\sim$281M parameters), we evaluate reduced variants with 12 blocks ($\sim$141M parameters) and 6 blocks ($\sim$94.4M parameters). All models are trained for 100k steps under identical training settings.

Table~\ref{table:exp3} reports the objective and subjective evaluation results. From the objective metrics, a gradual degradation can be observed as the model size decreases. Notably, while the 12-block and 6-block variants exhibit higher FD and KL values than the 24-block model, their CLAP scores remain competitive, suggesting that semantic alignment between text and audio is largely preserved even in smaller configurations.

The subjective evaluation, averaged over 19 participants, shows a consistent trend: perceptual generation quality generally decreases as the number of LM blocks is reduced. Nevertheless, despite its compact size, the 6-block Prefix\,SiMBA/Diffusion model still achieves reasonable perceptual quality across all three subjective criteria, demonstrating that the proposed SSM-based language model remains effective under parameter-constrained settings.

Overall, these results indicate that Prefix,SiMBA/Diffusion exhibits favorable parameter efficiency, enabling competitive text-to-music generation quality even when the model size is substantially reduced.

\section{Conclusion}

In this paper, we have systematically investigated the potential of SSMs for text-to-music generation. Our study leads to three key findings. First, we demonstrate that Prefix\,SiMBA, a selective SSM-based architecture, performs significantly better than its Transformer counterpart under similar training resources. Second, we show that the proposed two-stage Prefix\,SiMBA/Diffusion model compares favorably with the official MusicGen-small, despite using far fewer training resources.
Finally, our results indicate that Prefix,SiMBA maintains competitive text-to-music generation quality even with a substantially reduced number of parameters, demonstrating favorable parameter efficiency.
These results suggest that SSM-based architectures, particularly those integrating SSM and channel mixing like SiMBA, offer superior training efficiency compared to standard Transformers in resource-constrained TTM scenarios.
Future work could explore scaling up the SSMs to challenge larger SOTA models, and replacing  Transformers by SSMs 
in diffusion-Transformer architectures (DiTs) for TTM, such as Stable Audio Open~\cite{evans2025stable}.

\bibliography{IEEEexample}

\end{document}